\documentclass{iau-JDSS}
\usepackage{graphicx}

\title[short title of paper] 
{Are pre-MS stars older than we thought?}

\author[short author list]   
{Tim Naylor \& N.J. Mayne}

\affiliation{School of Physics, University of Exeter, EX4 4QL, UK}

\pubyear{2009}
\volume{Volume 15}  
\pagerange{119--126}
\date{?? and in revised form ??}
\jname{Highlights of Astronomy, Volume 14}
\editors{Ian F Corbett, ed.}
\begin{document}

\maketitle

\begin{abstract}
We present a consistent age ordering for young clusters and groups 
determined using the contraction of stars through their pre-main-sequence 
phase.
We compare these with ages derived from the evolution of the upper 
main-sequence stars, and find the upper MS ages are older by a factor 1.5 to 2.
We show that increasing the binary fraction and number of equal-mass binaries
amongst the O-stars compared to the rest of the MS cannot remove this 
discrepancy.

\end{abstract}

\begin{table}\def~{\hphantom{0}}
  \begin{center}
  \caption{Ages from PMS contraction}
  \label{tab:kd}
  \begin{tabular}{ccccccccccc}\hline
  1Myr  & 2Myr & 3Myr &     4-5Myr  &     & 5-10Myr          & 10Myr     & 13Myr            &  40Myr  \\
\hline
 IC5146 & ONC   & $\lambda$ Ori, &  IC348      &     & $\gamma$ Vel$^2$ &  NGC7160 & h \& $\chi$ Per & NGC2547  \\
     & NGC 6530  & $\sigma$ Ori, &Cep OB3b$^1$  \\
     &          & NGC2264        & NGC2362      \\
\hline
  \end{tabular}
 \end{center}
$^1$ Littlefair \etal\ in prep.\ \ \ $^2$\cite{Jeffries09}
\end{table}

In \cite{Mayne07} and \cite{Mayne08} we developed an age-ordering for
young stars and groups based on the luminosity of the pre-main-sequence.
Table 1 shows the resulting ages, including those derived in subsequent papers.
In \cite{Naylor09} we derived ages by fitting the
change position in the colour-magnitude diagram of upper-MS stars
as they evolve from the zero-age MS to the terminal-age MS.
We found that the MS ages are a factor 1.5 to 2 longer than
the ages derived from the PMS.

After my presentation Gaspard Duch\^ene pointed out that binarism amongst 
O-stars is much higher than in the rest of the MS.
A higher binary fraction will shift the centroid of the combined
single-star and binary-star sequences redwards, mimicking an older age
and perhaps explaining the older MS ages.
As the mass-ratio distribution is equally important, we tested this
idea using the most extreme assumption we could reasonably make,
the strong hypothesis of \cite{lucy06}, which we approximated as 25\% of binaries evenly 
distributed over $0.95<q<1.0$, and 75\% evenly distributed 
over $0.2<q<0.95$.
Using this, and a binary fraction (restricted to $q>0.2$) of 75\% 
(e.g. \cite{Sana09}) for all O-stars
we find the ages of the clusters change by less than 5\% compared with
the results of \cite{Naylor09}.
Thus the discrepancy between the MS and PMS ages remains.

\end{document}